\documentclass{IEEEtran}
\usepackage{amsmath,amsthm}
\usepackage{authblk}
\usepackage{algorithmic}
\usepackage{algorithm}
\usepackage{stmaryrd}
\usepackage{pxfonts}
\usepackage{graphicx}
\usepackage{array}
\usepackage{booktabs}
\usepackage{flushend}
\usepackage{subfigure}

\title{Avoiding Whitewashing in Unstructured Peer-to-Peer Resource Sharing Network}
\author{Ruchir Gupta, Yatindra Nath Singh, \emph{Senior Member IEEE}}
\affil{Department of Electrical Engineering, IIT, Kanpur}
\affil{\textit {\{rgupta,ynsingh\}@iitk.ac.in}}

\begin {document}

\maketitle
\begin {abstract}
In peer-to-peer file sharing network, it is hard to distinguish between a legitimate newcomer and a whitewasher. This makes whitewashing a big problem in peer-to-peer networks. Although the problem of whitewashing can be solved using permanent identities, it may take away the right of anonymity for users. In this paper, we a have proposed a novel algorithm to avoid this problem when network uses free temporary identities. In this algorithm, the initial reputation is adjusted according to the level of whitewashing in the network.
\end {abstract}
\section{Introduction}
With the tremendous growth in the number of Internet users, the traditional client-server systems are proved to be less scalable. This is because load on a server increases with the increase in the number of clients. The clients or user machines have become more powerful in the recent past. These two things pushed the evolution of peer-to-peer computing. Peer-to-peer computing is a type of computing where a peer or user machines act as server as well as client. There is no central control or authority in the network. All peers are equal in status. Peer-to-peer networks are typically used for resource sharing within the network. Resource may be storage, processing power or some kind of file.

Distributed nature of peer-to-peer networks poses many challenges. While designing protocols for peers, systems designers usually assume that peers in the network are simple machines and peers will not play games by modifying the protocols i.e., every peer will behave in selfless and hence co-operative manner.

But nodes are the entities operated by rational human beings. So nodes are expected to behave in a selfish manner i.e. they try to maximise their utility. This results in their non co-operative behaviour. This phenomenon is explained by famous Prisoners' Dilemma, in which Nash Equilibrium (NE) is achieved when both prisoners deceive each other \cite{osborne}. Similarly in a file sharing network, if peers are considered as players, their NE happens when none of them share the resources \cite{Tang}. Tendency of nodes to draw resources from the network and not giving any thing in return is termed as 'Free Riding'. 

Experimental studies on Gnutella network, first in 2000 \cite{Adar00freeriding}, second \cite{Saroiu02ameasurement} in 2002 states that more than half the number of total files shared in the network are from just 7\% or less number of peers where as 25\% or more number of peers share no file. Third study in 2005 \cite{Hughes2005} also confirmed this fact by showing that number of free rider nodes is as high as 85\%.

Different incentive or disincentive mechanisms are generally used to avid free riding. In these mechanisms, incentive is given to a peer if it co-operate and disincentive is given if it does not co-operates. This incentive is generally better quality of service or resource allocation. A reputation management system is required to manage these incentives and disincentives. As peer-to-peer networks are totally decentralised systems, we need a distributed reputation management system. Such kind of reputation management systems has been suggested by the authors in \cite{ruchir}. 

When these mechanisms are implemented, selfish users start whitewashing and colluding \cite{Feldman2004}. When a node has bad reputation in the network then to avoid disincentives, it leaves the network and returns back with a new identity as a new comer to the network. This is termed as whitewashing. Whitewashing problem becomes more challenging as it is difficult to differentiate between a legitimate newcomer and a whitewasher.

In this paper we analyses the pay-off and penalty for cooperative and defector node respectively in a reputation management system. We also propose an adaptive algorithm to counter the whitewashing attack.

Remainder of this paper is organised as follows. Section  \ref{RW4} discusses the
related work. Section \ref{systemmodel} describes the system model. Section \ref{PRG4} computes the pay-off in the reputation game. In section \ref{PW4}, the problem of whitewashing is discussed along with an algorithm as its solution and its analysis, Section \ref{NR4} presents the numerical results and section \ref{CON4} concludes the paper.
\section{Related Work}
\label{RW4}
Many authors have suggested that whitewashing can be totally removed if system has permanent identities \cite{castro,Izhak-Ratzin:2012}. Problem of whitewashing due to availability of cheap or free identities was initially pointed out by Friedaman \emph{et.al.} \cite{Friedman00}. They proved that considering every newcomer as defector is a better policy than any other static stranger policy. Whereas Feldman \emph{et.al.} \cite{Feldman2004_1} proved that newcomers should only be punished if turnover rate is high. They also proved that a legitimate user can only win from a whitewashing user if newcomer will be served well with a very small probability \cite{Feldman2005}. Yang \emph{et.al.} studied Maze file sharing system and concluded that incentives promote whitewashing \cite{Yang:2005}.

Lai \emph{et.al.} suggested in \cite{Feldman2004_1,Lai03} that newcomers should be served as per the behaviour shown by them at that time. But this is more prone to collusion and moreover in a resource sharing network, nodes generally form a group where they serve more. In that case observing behaviour will not work well. Similar kind of approaches are proposed for new service providers' reputation among consumers in on-line communities \cite{Malik2009}.  

Anceaume \emph{et.al.} \cite{Anceaume:2005} have suggested to charge an entry fee from every newcomer in such a way that it should avoid whitewashing and should not discourage newcomers. Chen \emph{et.al.} claims that a whitewashing node will continue to have same kind of habits even after whitewash so it should be identified and punished on the basis of its habits \cite{Chen:2009}. Zuo \emph{et.al.} suggested that newcomers should only be allowed to access the resource from newcomers and low trust value nodes \cite{multilablel}.

Yu \emph{et.al.} \cite{whitewashaware} proposed that any user that have served even once should have a higher value of reputation then a newcomer and it should rise faster initially and slower later on as a result of cooperation.

\section{System Model}
\label{systemmodel}
In this paper, we are studying a pure unstructured peer-to-peer network. Peers in this network are connected by an access link followed by a backbone link and then again by an access link to the second node. We are assuming that the network is heavily loaded i.e. every peer has sufficient number of pending download requests, hence these peers are contending for the available transmission capacity. We also assume that every peer is paying the cost of access link as per the usage. So, every peer wants to maximises its download and minimise its upload so that it can get maximum utility for its money spend in access link.

Considering nodes to be purely rational does not answer many questions about nodes' behaviour \cite{Feldman2004_1}. Feldman \emph{et.al.} \cite{Feldman2004_1} proposed a model where rationality depend upon the type of node or the level of generosity i.e. every node will free ride or contribute as per its type. On the similar lines, we propose the level of honesty regarding whitewashing behaviour of nodes. If $h_i$ be the honesty level of $i^{th}$ node then $i$ will whitewash provided,
 \begin{equation}
 R_{ini}\ge h_i
 \end{equation}
 Here $R_{ini}$ is the initial reputation given to a newcomer node.
It means that every node is a potential white washer as per its level of honesty. If pay-off by whitewash is greater than its honesty level, then it will whitewash otherwise it will remain an honest node. As network will contain all kind of nodes with equal probability, honesty level of nodes are assumed to be distributed uniformly between 0 to 1.
 
Reputation can be measured in number of ways. We measure the reputation of a node after a transaction as,
\begin{equation}
\label{1}
Reputation\; of\; node\; i= \frac{Resource \; provided\; by\;node\; i}{Resource \; requestd\; to\; node\;i}.
\end{equation}
\section{Pay-off in Reputation Game}
\label{PRG4}
In this section, we will see the pay-off of an agent in a network where a reputation management system has been implemented with different type of identities viz. permanent identities and temporary identities. Permanent identities can also be termed as $\infty$ cost identities as these can not be changed. Social security number in USA is one such example. Whereas temporary identities can be classified into two types, finite cost identities and zero cost identities on the basis of cost of assigning a new identity. 

Here it is important to note that in an interaction even if cost incurred in serving other nodes' request and the value received from other nodes in response to its request, are same, the node still gains by a small amount $\delta$. This comes because after every transaction the chance of survival satisfaction for a node increases.

\subsection{Pay-off of agents with permanent identities}
Let us assume that there are $N$ nodes in the network. Reputation of these nodes is distributed between $0$ and $1$ with some arbitrary distribution having mean $\mu$. When a new node joins the network, it is assigned an initial reputation value of $R_{ini}$. We are assuming that reputation aggregation is happening after every round of transactions and in this process, the nodes that have transacted with the node will jointly form a reputation that will be spread across the network. It means that every node knows the reputation of other node after every round. For simplicity we assume that every node requests for same amount of resource i.e. $c$. We also assume that network is large and hence the mean reputation of the network will not change considerably and probability of allocation against a node's request will be
\begin{equation}
\nonumber P(allocation)\propto (reputation)^x 
\end{equation}
Taking proportionality constant to be $1$ for simplicity
\begin{equation}
P(allocation)= (reputation)^x.
\end{equation}
As we have stated, the expected reputation of nodes will be $\mu$. Any node can ask for a resource from the entering node. The probability that a particular node will ask for a resource is $\frac{1}{N}$ assuming that all nodes are equally likely to ask for a resource. Let us assume that every node can ask $c$ amount of resource. So, if the node receives one request, the expected service, it has to do, will be
\begin{eqnarray}
\nonumber E[Serv(1)]&=& \frac{1}{N}\cdot\mu^x\cdot c+\frac{1}{N}\cdot\mu^x\cdot c+...+\frac{1}{N}\cdot\mu^x\cdot c\\
&=& \mu^x\cdot c.
\end{eqnarray}
Whereas the node will get service according to its reputation. So the expected return for this node if it makes a single request will be
\begin{equation}
E[ret(1)]=c\cdot (R_{ini})^x+\delta.
\end{equation}
First we will see the pay-off of cooperative node. If a node serves $m$ requests and makes $\acute{m}$ requests,  the expected  pay-off of this node at the end of the first round will be
\begin{equation}
E_c[payoff(1)]=-m\mu^x\cdot c+\acute{m}c\cdot (R_{ini})^x+\delta.
\end{equation}
The expected reputation of an entering node after first round of transaction, if reputation is measured using equation (\ref{1}), is
\begin{eqnarray}
\nonumber E_c[R(1)]&=&\frac{m\mu^x\cdot c}{mc},\\
&=& \mu^x.
\end{eqnarray}
The reputation of a node $i$ is estimated by few other nodes. The nodes with whom interaction does not take place use $R_{ini}$ as the default reputation. When aggregating the reputation \cite{ruchir}, only the estimated reputations are aggregated, not the default value $R_{ini}$.

 Similarly pay-off and reputation of this node at the end of second round will be,
\begin{eqnarray}
\nonumber E_c[payoff(2)] &=&-2m\mu^x\cdot c+\acute{m}\cdot(\mu^x)^x\cdot c\\\nonumber&&+\acute{m}\cdot (R_{ini})^x c+2\delta\\
\nonumber E_c[R(2)]&=& \frac{\mu^x+\mu^x}{2}=\mu^x
\end{eqnarray}
Similarly, pay-off and reputation of this node at the end of $k^{th}$ round will be
\begin{eqnarray}
\label{coppay}
 E_c[payoff(k)] &=&-km\mu^x \cdot c+(k-1)\acute{m}\cdot(\mu^x)^x\cdot c\\\nonumber&&+\acute{m}\cdot (R_{ini})^x c+k\delta\\
 E_c[R(k)]&=& \frac{\mu^x+...+\mu^x}{k}=\mu^x
\end{eqnarray}
Now let's consider defecting node. In the first round it will get service as per $R_{ini}$. But after that, it will not get any service because every body will know that the node with this identity is not going to serve. It can not change its permanent identity as this will cost him $\infty$. So the expected pay-off of defecting node, with same request profile, after $k$ rounds will be
\begin{equation}
\label{difpay}
E_d[payoff(k)]=\acute{m}\cdot (R_{ini})^xc+\delta.
\end{equation} 
If we take $m=\acute{m}$, as this should be the condition statistically, it can be seen that for $x\ge 1$, cooperative node will always remain in negative pay-off if it has a $R_{ini}$ less then $\mu$. Moreover for any value of $R_{ini}$ with $x\ge 1$, the defector node will always remain in advantage. So value of $x$ should be less then $1$. Now let's compare (\ref{coppay}) and (\ref{difpay}) to get the number of rounds required after which cooperative node will be in advantage over defector node.
\begin{equation}
\nonumber \acute{m}\cdot (R_{ini})^xc +\delta\le -km\mu^x \cdot c+(k-1)\acute{m}\cdot(\mu^x)^x\cdot c+\acute{m}\cdot (R_{ini})^x c+k\delta
\end{equation}
On solving, we get
\begin{equation}
k\ge\frac{\acute{m}(\mu^x)^x+\frac{k\delta}{c}}{\acute{m}(\mu^x)^x+\frac{k\delta}{c}-m\mu^x}.
\end{equation}
Considering $\delta$ to be small and minimizing the value of $k$ with respect to $x$, it turns out that for $x=\frac{1}{2}$, $k$ will be minimum. So for $x=\frac{1}{2}$, $k$ will be
\begin{equation}
k\ge\frac{\acute{m}\mu^\frac{1}{4}}{\acute{m}\mu^\frac{1}{4}-m\mu^\frac{1}{2}}
\end{equation}
Here it is interesting to note that $\acute{m}=m$ (that should be the condition under equilibrium) and $\mu=1$ value of $k$ turns out to be $\infty$. This happens because now newcomer node can not improve its reputation any more above $\mu=1$ and if node is cooperative, every time its positive pay-off becomes equal to its negative pay-off. Only $\delta$ gain is the gain that comes as an advantage to him over the defector. Secondly if every body has reputation equal to $1$ that means it's an ideal system and does not need any reputation management. 
\subsection{Pay-off of agents with zero cost identities}
\begin{figure}[!t]
\begin{center}
\includegraphics[width=88.5mm, height=85mm, keepaspectratio=false]{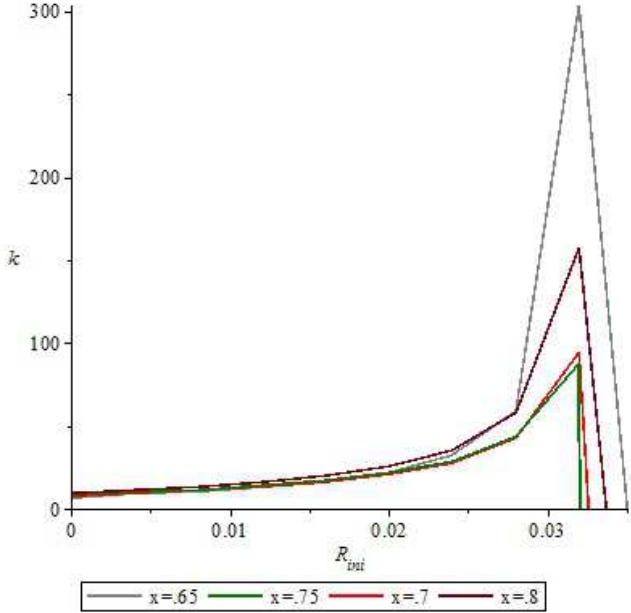}
\caption{Variation in Initial reputation, number of round and x}
\label{rkxf}
\end{center}
\end{figure}

In case of free identities, the pay-off for a cooperative node will remain same as in the case of permanent identities i.e. (\ref{coppay}). But for defector node, it will change because now defector node can white wash and change its identity. So the defector's expected pay-off after first round will be,
\begin{equation}
\nonumber E_w[payoff(1)]=\acute{m}\cdot c\cdot (R_{ini})^x+\delta.
\end{equation}
Its reputation will become $0$ as it will not serve but it will change its identity as it will not cost any thing for it. So its reputation will again become $R_{ini}$.
Pay-off of defector after second round will be,
\begin{eqnarray}
\nonumber E_w[payoff(2)]&=&\acute{m}\cdot c\cdot (R_{ini})^x+\delta+\acute{m}\cdot c\cdot (R_{ini})^x+\delta\\\nonumber &=& 2\acute{m}\cdot c\cdot (R_{ini})^x+2\delta.
\end{eqnarray}
Similarly pay-off after $k$ rounds will be,
\begin{equation}
\label{wwp}
E_w[payoff(k)]=k\acute{m}\cdot c\cdot (R_{ini})^x+k\delta.
\end{equation}
Comparing (\ref{coppay}) and (\ref{wwp}) to get the number of rounds required after which cooperative node will be on advantage over defector node.
\begin{equation}
\nonumber k\acute{m}\cdot (R_{ini})^xc+k\delta \le -km\mu^x \cdot c+(k-1)\acute{m}\cdot(\mu^x)^x\cdot c+\acute{m}\cdot (R_{ini})^x c+k\delta
\end{equation}
On solving we get,
\begin{equation}
k\ge\frac{\acute{m}(\mu^x)^x-\acute{m}(R_{ini})^x}{\acute{m}(\mu^x)^x-m\mu^x-\acute{m}(R_{ini})^x}.
\end{equation}
For $\mu=0.5$ and $m=\acute{m}$, it turns out that the maximum possible value of $R_{ini}$ giving which ensures that a cooperative node will win against a defector node  is $0.036$ at $x=0.70$ (figure \ref{rkxf}). 

\subsection{Pay-off of agents with finite cost identities}
Now we will consider an intermediate case where identities have a fixed finite cost, i.e. $z$. It means that every time when a node will enter in the network with a new identity, it will get a pay-off of $-z$. The expected  pay-off of the cooperative node in the end of first round will be
\begin{equation}
E_c[payoff(1)]=-m\mu^x\cdot c+\acute{m}c\cdot (R_{ini})^x-z+\delta.
\end{equation}
The expected reputation of entering node after first round of transaction if reputation is measured using equation (\ref{1}), is given by
\begin{eqnarray}
\nonumber E_c[R(1)]&=&\frac{m\mu^x\cdot c}{mc}\\
&=& \mu^x.
\end{eqnarray}
 Similarly pay-off and reputation of this node at the end of second round will be
\begin{eqnarray}
\nonumber E_c[payoff(2)] &=&-2m\mu^x\cdot c+\acute{m}\cdot(\mu^x)^x\cdot c\\\nonumber&&+\acute{m}\cdot (R_{ini})^x c-z+2\delta\\
\nonumber E_c[R(2)]&=& \frac{u^x+\mu^x}{2}=\mu^x
\end{eqnarray}
Similarly pay-off and reputation of this node at the end of $k^{th}$ round will be,
\begin{eqnarray}
\label{coppayf}
 E_c[payoff(k)] &=&-km\mu^x \cdot c+(k-1)\acute{m}\cdot(\mu^x)^x\cdot c\\\nonumber&&+\acute{m}\cdot (R_{ini})^x c-z+k\delta\\
 E_c[R(k)]&=& \frac{\mu^x+...+\mu^x}{k}=\mu^x.
\end{eqnarray}
The defector's expected pay-off after first round will be,
\begin{equation}
\nonumber E_w[payoff(1)]=\acute{m}\cdot c\cdot (R_{ini})^x-z+\delta.
\end{equation}
Its reputation will become $0$ as it will not serve but it will change its identity as it will not cost any thing for it. So its reputation will again become $R_{ini}$.
Pay-off of defector after second round will be,
\begin{eqnarray}
\nonumber E_w[payoff(2)]&=&\acute{m}\cdot c\cdot (R_{ini})^x-z+\acute{m}\cdot c\cdot (R_{ini})^x-z+2\delta\\\nonumber&=& 2\acute{m}\cdot c\cdot (R_{ini})^x-2z+2\delta.
\end{eqnarray}
Similarly pay-off after $k$ rounds will be,
\begin{equation}
\label{wwpf}
E_w[payoff(k)]=k\acute{m}\cdot c\cdot (R_{ini})^x-kz+k\delta.
\end{equation}
Comparing (\ref{coppayf}) and (\ref{wwpf}) to get the number of rounds required after which cooperative node will be on advantage over defector node, we get
\begin{equation}
\label{wwpf1}
 k\acute{m}\cdot c\cdot (R_{ini})^x-kz+k\delta\le k m\mu^x\cdot c+(k-1)\acute{m}\cdot(\mu^x)^x\cdot c+\acute{m}\cdot (R_{ini})^x c-z+k\delta.
\end{equation}
\begin{equation}
k\ge\frac{\acute{m}(\mu^x)^x-\acute{m}(R_{ini})^x+\frac{z}{c}}{\acute{m}(\mu^x)^x-m\mu^x-\acute{m}(R_{ini})^x+\frac{z}{c}}.
\end{equation}
\section{Problem of Whitewashing}
\label{PW4}
When in a peer-to-peer network, a reputation management system is implemented, selfish nodes start applying different strategies to counter the rules of the system for their benefit. Whitewashing is one such strategy.

If cheap identities are available in the system, a rational node does not cooperate i.e. it free rides and changes its identity when its reputation goes low. When a normal node encounters such node, it thinks that this is a legitimate newcomer and hence serves it according to that. Hence, a white washer utilizes the resources of system even without cooperating. 

Whitewashing can be totally avoided if we have permanent identities as mentioned in the previous section. But having such permanent identities implies that nobody is anonymous \cite{Friedman00}. One way to get permanent identities without losing anonymity is to use Gossip-Based Computation of Aggregate Information. This requires a trusted central certification authority and every user has to take a signed certificate.

If permanent identities are not to be used, there is no way by which one can differentiate between legitimate newcomer and a whitewasher. Whitewashing can be avoided by keeping a low initial reputation value $R_{ini}$. But as we have seen in the previous section, this leads to a very small value and discourages legitimate new comers to join the system because of initial hardship in the system. 

Keeping some cost of identities seems to be a good idea. But it is difficult to decide the cost of an identity. If the cost is kept higher, the whitewashing will be discouraged but on the other hand legitimate new comers will also be discouraged. This will reduce the efficiency of the system. Moreover, distribution of the earning due to the cost of identities among the existing users, will be another problem. Alternatively a policy may be made that any newcomer will not be served initially or will be served with poor quality of service. It will be served properly only once it has earned some credit by serving some existing nodes i.e. a newcomer will be provided small or $0$ initial reputation. This will lead to initial hardship to legitimate newcomers and hence will decrease system performance.

We propose that when a new node joins the network, it should be given an initial reputation value $R_{ini}$. As it's difficult to identify whether a new joining node is a a whitewasher or legitimate newcomer, all we can do is to make decision on the basis of level of whitewash in the network. So this initial value should keep on changing on the basis of level of whitewashing in the network. It means that if whitewashing level is low, the $R_{ini}$ value will be kept high; where as if whitewashing level increases, $R_{ini}$ value will be decreased adaptively.

As whitewashing nodes can not be identified and hence can not be counted so we need to estimate the level of whitewashing on the basis of growth in network at regular basis. For this, we need to follow an algorithm for estimation of whitewashing level. In this algorithm, we count the number of nodes at regular basis and finds the increase in the size of network in distributed fashion. This can be easily done using gossiping techniques. If reputation management system is based on \cite{ruchir}, gossiping is already happening at regular intervals. So there will not be any additional overhead.

As we are considering network based on PA model \cite{bollobas}, the newcomer node will join the network preferably attaching to the existing nodes with high degree. Hence, growth rate at different places within the network will be different. 

Once a node finds the total number of nodes in the network at two consecutive gossiping instances, it can locally calculate the overall growth rate of network. This estimate of growth rate must be adjusted to get an estimate of local growth rate due to the reason mentioned above. This can be done using,
\begin{equation}
G_{local,i}=\frac{d_{local,i}}{d_{average}}\cdot \frac{N_{n}}{N_{n-1}}.
\end{equation}
Here $G_{local,i}$ is local growth rate of the network with respect to node $i$; $d_{average}$  is the average degree of complete network; $d_{local,i}$ is the local average degree with respect to node $i$; and $N_{n}$, $N_{n-1}$ are the network size in current and previous iteration respectively. Overall average degree of the network can be found out using gossip. By adding one more parameter, i.e. self degree of node, to gossiping will give the sum of degrees of nodes in the network and hence average degree can be estimated by every node. As gossip is already happening, adding new parameter in the gossip will not lead to too much increase in overhead. The local average degree can be computed by taking average of neighbours' degree. 

The nodes will also monitor see the recently departed nodes. It will categorize theses into two types on the basis of Expected initial reputation. Expected initial reputation is the average of maximum initial reputation and minimum initial reputation. First type of nodes have higher or equal reputation than the expected initial reputation. These nodes will be considered as legitimate departing nodes as no node will whitewash if it already has a reputation higher than the expected initial reputation. Second type of nodes are with lower reputation then the expected initial reputation. These will be considered as potential whitewasher nodes. A node shares following information about its neighbourhood - number of nodes at the end of previous iteration, number of departing nodes and newly arriving nodes since last iteration. If there is no whitewashing, then newly arriving nodes ($A_i$) should be equal to sum of legitimate departing nodes ($L_i$) and expected localized growth, i.e.,
\begin{equation}
\nonumber A_i=L_i+N_{i,n-1}\cdot (G_{local,i}-1).
\end{equation}
If whitewashing happens, then
\begin{equation}
\nonumber A_i>L_i+N_{i,n-1}\cdot (G_{local,i}-1).
\end{equation}
The difference is expected to b number of whitewashers $\acute{w}_i$. Thus
\begin{equation}
\nonumber A_i-\acute{w}_i=L_i+N_{i,n-1}\cdot (G_{local,i}-1).
\end{equation}
 We can define the level of whitewashing $W_{i,n}$ by node $i$ as
\begin{equation}
W_{i,n}=\frac{\sum\limits_{j\in Neg(i)}A_{j}-N_{j,n-1}\cdot (G_{local,j}-1)-L_{j}}{\sum\limits_{j\in Neg(i)}N_{j,n}}.
\end{equation}
In this formula, $N_{i,n}$ is the sum of degree of the neighbours of a node at the current iteration and $Neg(i)$ is the set of neighbours of node $i$.

We assume that node will vary the value of initial reputation with respect to level of whitewashing in quadratic fashion. 
Let $R_{ini,i,max}$, $R_{ini,i,n}$ and $R_{ini,i,min}$ be the initial reputation values at zero whitewashing, current whitewashing level ($W_{i,n}$) and maximum possible whitewashing level at node $i$ respectively.
We propose the initial reputation $R_{ini,i,n}$ as,
\begin{equation}
\label{23}
R_{ini,i,n}=\begin{cases}\tilde{R}_{ini,i,n} &\text{if $\tilde{R}_{ini,i,n}\ge R_{ini,i,min}$}. \\
	 R_{ini,i,min}, &\text{otherwise}.
	\end{cases}
	%
\end{equation}
where
\begin{equation}
\nonumber \tilde{R}_{ini,i,n}	=\left(1-\frac{W_{i,n}}{W_{i,max}}\right)^2\cdot R_{ini,i,max}
\end{equation}
The $W_{i,max}$ is the maximum possible whitewashing level at the node $i$. Initially it will be taken as the ratio of number of nodes that have honesty level between $0$ and $R_{ini,n,max}$ and total nodes. Hence, it turns out to be $(R_{ini,n,max}\cdot N)/N=R_{ini,n,max}$. Later on node will try to estimate it by taking the maximum whitewashing level seen over last $\acute{n}$ rounds such that
\begin{equation}
W_{i,max}=\max_{last\; \acute{n}\; rounds}W_{i,n}.
\end{equation} The value of $\acute{n}$ may be dynamically adjusted by the node as per the change observed in whitewashing level. For this paper, $\acute{n}$ has been taken as constant for every node for simplicity. The value of this constant has been taken as 10 for simulation

The value of $R_{ini,i,min}$ will be taken such that a cooperative users may win over whitewashing users after few rounds. As evident from the figure \ref{rkxf}, number of rounds required to win a cooperative user increases at a very high rate above a certain $R_{ini,i}$ value. It may be taken as $R_{ini,i,min}$.

 The value of $R_{ini,i,max}$ will be decided on the basis of the cooperation received from legitimate newcomers. A node will average the reputations of newcomers nodes after every gossiping round. In this average computation, it will consider the nodes that have seen three gossiping rounds. This average may be taken as $R_{ini,i,max}$.
 
\begin{figure*}[!t]
\begin{center}
\includegraphics[width=140mm, height=80mm, keepaspectratio=false]{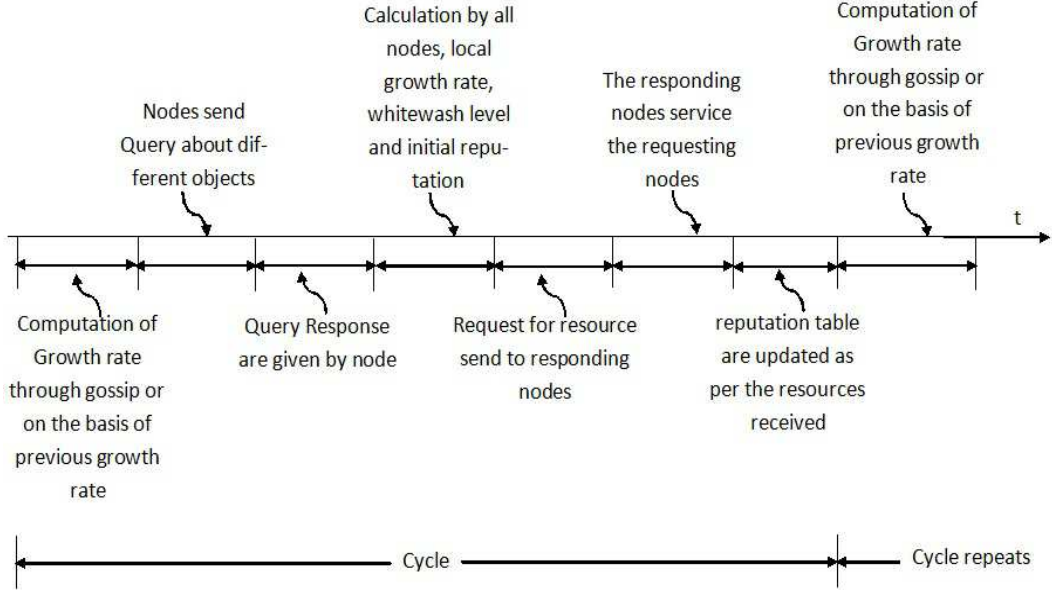}
\caption{Timing Diagram}
\label{timing}
\end{center}
\end{figure*}

 Timing diagram of whole process that a node will undergo is described in figure \ref{timing}. We can observe the following from this figure. Any node will periodically allocate resource to the requesting nodes. In this process, a node will first reply to the received queries, then on the basis of number of queries received and network growth rate, it will estimate the level of whitewashing. After this estimation, node will allocate resources to the requesting nodes.  The new nodes that will join the network after whitewash level estimation will be served resources after the next whitewash level estimate. Nodes can estimate the growth rate by interpolating the previous gossips until the next round of gossip is completed.

\subsection{Analysis of Algorithm}
\label{analysis}
While analysing the algorithm, we need to understand that a node has two choices, first, to be cooperative and the second to be non-cooperative. While being non-cooperative, a node have the choice to whitewash to maximize his gain. Usually non cooperative users will whitewash as that is the only way it can gain from the system. A user will be non-cooperative and hence will also whitewash if its honesty level is lesser than the current initial reputation allotted to any newcomer node. As the current initial reputation level will always lie in between the maximum initial reputation $R_{ini,i,max}$ and minimum initial reputation $R_{ini,i,min}$, we can state the following. All nodes having honesty level higher then $R_{ini,i,max}$, will never whitewash and thus will be always be cooperative. All nodes having honesty level lesser than $R_{ini,i,min}$ will always whitewash. All the nodes having honesty level between $R_{ini,i,max}$ and $R_{ini,i,min}$ will choose to be either cooperative or non cooperative (thus be whitewasher) depending on expected current value of initial reputation.

If in the system, the current value of initial reputation is maintained such that the nodes who are whitewashing in every round will certainly loose to cooperative nodes, the nodes will have no choice but to be cooperative.

The nodes who choose to be non-cooperative, have a choice of whitewashing or not. In case whitewashing nodes are less, the whitewashing level will be less and thus $R_{ini,i,n}$ will be set to higher value and vice-versa. Let us understand, what should be natural choice of non-cooperative node - to whitewash or not to whitewash.
 
To understand the process that is happening with the nodes, we will start with the simple case when only two players have their honesty level below the maximum initial reputation and above the minimum initial reputation. There are three possibilities in this case. First, no node whitewashes, second, one of the nodes whitewash and third when both nodes whitewash. Here, the maximum whitewash level is $\frac{2}{N}$. Hence, in first case $R_{ini,i,n}$ will be $R_{ini,i,max}$, in second case, it will be $\frac{R_{ini,i,max}}{4}$ and in third case, it will be $R_{ini,i,min}$ according to \eqref{23}. Therefore the game matrix will be as follows,
 \begin{table}
 \begin{center}
  \begin{tabular}{c|c|c|c|}
 \multicolumn{4}{c}{B}\\
 \cline{2-4}  && Do whitewash & Don't whitewash \\ 
 \cline{2-4} A & Do whitewash  &($R_{ini,i,min}$, $R_{ini,i,min}$)  & ($\acute{R}$, 0) \\ 
 \cline{2-4} & Don't whitewash & (0, $\acute{R}$) & (0, 0) \\ 
 \cline{2-4} 
  \end{tabular} 
 \end{center}
  \caption{Payoff of a whitewasher - pure strategy}
   
   \end{table}
Here $\acute{R}=\frac{R_{ini,i,max}}{4}$. By this matrix, it is evident that \{Do whitewash, Do whitewash\} is the weakly dominant strategy and hence both players will enter in every round and will get a payoff of  $R_{ini,i,min}$.   
   
 $R_{ini,i,min}$ is a payoff that is too low to whitewash as explained earlier, hence it remains no more lucrative option for such users. Consequently they do not whitewash and their payoff becomes zero.
  
This zero payoff discourages their non-cooperative, whitewashing tendency and consequently forces them for cooperation. Similar  thing will occur in $\kappa$ player game, i.e. when $\kappa$ players have their honesty level below the maximum initial reputation. In that case, game will be as follows.\\ \\
Player set P=\{nodes with honesty level below maximum initial reputation\}   \\ \\
Actions$\in$\{$S_1, S_2,...,S_\kappa$\}\\ 
here $S_i$=\{Do whitewash, Don't whitewash\} $\forall i\in \{1,2,3,...,\kappa\}$\\ \\

\begin{eqnarray}
\nonumber Payoff&=& u_i((v_1,...,v_{\kappa}))\\\nonumber&&=\left(1-\frac{\sum\limits_j v_j }{\kappa}\right)^2\cdot R_{ini,n,max} \text{if $R_{ini,i,n}\ge h_i$}. 
\end{eqnarray}
Nodes will not whitewash if $R_{ini,n} < h_i$.
Here $u_i$ is the pay-off of $i^{th}$ player and $v_j$ is the action performed by $j^{th}$ player such that, $v_j=1$ if player do the whitewash and 0 otherwise.
In this case also, it is evident that the weakly dominant strategy for all the  players is to whitewash. This leads to a payoff of $R_{ini,i,min}$. As said earlier $R_{ini,i,min}$ is a payoff that is too low to whitewash and it remains no more lucrative option for these users. Consequently they do not whitewash and their payoff becomes zero. This zero payoff discourages their non-cooperative, whitewashing tendency and consequently forces them for cooperation.

Users may also randomize their whitewashing action in each period, i.e. they may whitewash in a period with a particular probability. It is easy to observe that if there are $\kappa$ players then they will randomized over utmost $\kappa$ rounds  to maximize their randomization benefit. Let's examine a two player case when both are randomizing over two periods. The game matrix will be as follows.
\begin{table}
\begin{center}

\begin{tabular}{c|c|c|c|}
\multicolumn{4}{c}{B}\\
\cline{2-4}  & &  Round 1, (p) &  Round 2, (1-p) \\ 
 \cline{2-4} A &  Round 1, (q) & (0,0) & ($\acute{R}, \acute{R}$) \\ 
\cline{2-4}   &Round 2, (1-q) & ($\acute{R}, \acute{R}$) & (0,0) \\ 
\cline{2-4} 
\end{tabular} 
\end{center}
\caption{Payoff of a whitewasher - mixed strategy}
\end{table}
 Here, when both the players are whitewashing simultaneously, their payoff will become $R_{ini,i,min}$ and that will become zero because of reason that both the nodes will find whitewashing non-lucrative and will decide not to whitewash. Let us compute the payoff for player A for whitewashing in round 1 when player B is whitewashing in round 1 and round 2 with probabilities $p$ and $1-p$ respectively,
 \begin{equation}
 payoff(A,1)=p\cdot 0+(1-p)\cdot\acute{R}.
 \end{equation}
 Similarly for entering in round 2,
 \begin{equation}
 payoff(A,2)=(p)\cdot\acute{R}+ (1-p)\cdot 0.
 \end{equation}
 As player A is randomizing, pay-off in both the round should be same, so on equating both the equation we get $p=\frac{1}{2}$ and in similar fashion $q=\frac{1}{2}$. So the equilibrium probability distribution for p and q both is $\{\frac{1}{2},\frac{1}{2}\}$.
 
 Now let us examine a three player (A,B,C) case. There are two possibilities -- when players are randomizing over two rounds and when players are randomizing for three rounds. Let us assume that, the initial reputations given in some round when 1, 2 or 3 players are entering are $\acute{R_1}$, $\acute{R_2}$ and $\acute{R_3}$ respectively such that $\acute{R_1}> \acute{R_2} >\acute{R_3}$. The exact values of $\acute{R_1}$, $\acute{R_2}$ and $\acute{R_3}$ can be calculated using (\ref{23}),
 \begin{subequations}
 \label{r1r2r3}
 \begin{align}
  \acute{R_1}&=\left(1-\frac{1}{3}\right)^2\cdot R_{ini,i,max}=\frac{4}{9}R_{ini,i,max}\\
 \acute{R_2}&=\left(1-\frac{2}{3}\right)^2\cdot R_{ini,i,max}=\frac{1}{9}R_{ini,i,max}\\
  \acute{R_3}&=\left(1-\frac{3}{3}\right)^2\cdot R_{ini,i,max}=R_{ini,i,min} 
  \end{align}
 \end{subequations}
  As we know that players have different level of honesty, we assume that A has highest level of honesty and will only whitewash for $\acute{R_1}$, B will whitewash for $\acute{R_1}$ and $\acute{R_2}$ and C will whitewash for all values of initial reputation.
 
 Lets us first consider randomization over three rounds. Probabilities for whitewashing in round 1, 2 and 3 are $p_A, q_A, r_A; p_B,q_B,r_B$ and $p_C,q_C, r_C$ for players A,B and C respectively.
 
 Equating the pay-off of player A for whitewashing in round 1, 2 or 3,
 \begin{eqnarray}
(1-p_B)(1-p_C)\acute{R_1}=(1-q_B)(1-q_C)\acute{R_1}=(1-r_B)(1-r_C)\acute{R_1}
 \end{eqnarray}
 Here it may be noted that player A will only whitewash when neither B nor C are whitewashing.
 Similarly for player B
\begin{eqnarray}
 &&\nonumber(1-p_A)p_C\acute{R_2}+(1-p_C)p_A\acute{R_2}+(1-p_A)(1-p_C)\acute{R_1}=\\\nonumber&&(1-q_A)q_C\acute{R_2}+(1-q_C)q_A\acute{R_2}+(1-q_A)(1-q_C)\acute{R_1}=\\&&(1-r_A)r_C\acute{R_2}+(1-r_C)r_A\acute{R_2}+(1-r_A)(1-r_C)\acute{R_1}
\end{eqnarray}
Here it may be noted that player B will only whitewash when only one more or no player is whitewashing.
  and for player C
\begin{eqnarray}
  &&\nonumber\acute{R_3}p_Bp_A+(1-p_B)p_A\acute{R_2}+(1-p_A)p_B\acute{R_2}+\\\nonumber&&(1-p_B)(1-p_A)\acute{R_1}\nonumber=\acute{R_3}q_Bq_A+(1-q_B)q_A\acute{R_2}+\\\nonumber&&(1-q_A)q_B\acute{R_2}+(1-q_B)(1-q_A)\acute{R_1}=\acute{R_3}r_Br_A+\\&&(1-r_B)r_A\acute{R_2}+(1-r_A)r_B\acute{R_2}+(1-r_B)(1-r_A)\acute{R_1}
\end{eqnarray}
Here it may be noted that player C will whitewash in all conditions.
   And
\begin{equation}
   p_A+q_A+r_A=p_B+q_B+r_B=p_C+q_C+r_C=1
\end{equation}
Solving above set of equations we get $p_A=p_B=p_C=q_A=q_B=q_C=r_A=r_B=r_C=\frac{1}{3}$.
Hence, when randomizing over 3 rounds,the payoff of player A will be ($\frac{4}{9}\acute{R}_1$), for player B the payoff will be ($\frac{4}{9}\acute{R}_1+\frac{4}{9}\acute{R}_2$), and for player C the payoff will be ($\frac{4}{9}\acute{R}_1+\frac{2}{9}\acute{R}_2+\frac{1}{9}\acute{R}_2$). Using equation (\ref{r1r2r3}), the value of payoff for A, B and C will be ($\frac{16}{81}R_{ini,i,max}$), ($\frac{20}{81}R_{ini,i,max}$) and ($\frac{18}{81}R_{ini,i,max}+\frac{1}{9}R_{ini,i,min}$) respectively.

 Similarly if these three players will randomize over two periods with probabilities $p_A,q_A,p_B,q_B,p_C,q_C$ respectively, the probabilities will turn out to be $\frac{1}{2}$. Hence, when randomizing over 2 rounds, the payoff of player A will be ($\frac{1}{4}\acute{R}_1$), for player B the payoff will be ($\frac{1}{4}\acute{R}_1+\frac{1}{2}\acute{R}_2$) and for player C the payoff will be ($\frac{1}{4}\acute{R}_1+\frac{1}{2}\acute{R}_2+\frac{1}{4}\acute{R}_2$). Using equation (\ref{r1r2r3}), the value of payoff for A, B and C will be ($\frac{4}{36}R_{ini,i,max}$), ($\frac{6}{36}R_{ini,i,max}$) and  ($\frac{6}{36}R_{ini,i,max}+\frac{1}{4}R_{ini,i,min}$) respectively. 
 
It is evident from payoffs shown above that players will prefer to randomize over three rounds and not less then three. 

Now let us examine for $\kappa$ players ($A_1,A_2,...,A_{\kappa}$) randomizing over $\acute{\kappa}$ rounds. Let us assume that, the initial reputations given in some round when 1 to $\kappa$ players are entering are $\acute{R_1}$ to $\acute{R_{\kappa}}$ respectively such that $\acute{R_1}>\acute{R_2}>...> \acute{R_{\kappa}}$. The exact values of $\acute{R_1}$, $\acute{R_2}$ etc. can be calculated using (\ref{23}). As we know that players have different level of honesty, we assume that $A_1$ has highest level of honesty and will only accept $\acute{R_1}$, and $A_{\kappa}$ will accept all values of initial reputation.

 Let us assume that probability for whitewashing in $i^{th}$ round by $A_j$ is $p_{ji}$. Equating the pay-off of player $A_1$ for entering in all $\acute{\kappa}$ rounds,
 \begin{eqnarray}
 \label{31}
\nonumber&&(1-p_{21})(1-p_{31})...(1-p_{\kappa 1})\cdot \acute{R_1}\\\nonumber&=&(1-p_{22})(1-p_{32})...(1-p_{\kappa 2})\cdot \acute{R_1}\\&=&...=(1-p_{2\acute{\kappa}})(1-p_{3\acute{\kappa}})...(1-p_{\kappa\acute{\kappa}})\cdot \acute{R_1}
 \end{eqnarray}
 Similarly for player $A_2$
  \begin{eqnarray}
  \label{32}
 \nonumber&& \acute{R_2}\cdot[(p_{11})(1-p_{31})...(1-p_{\kappa 1})+(1-p_{11})(p_{31})...(1-p_{\kappa 1})\\\nonumber &&+...+(1-p_{11})(1-p_{31})...(p_{\kappa 1})]+\acute{R_1}\cdot(1-p_{11})(1-p_{31})\\\nonumber&&...(1-p_{\kappa 1})\\\nonumber&=& \acute{R_2}\cdot[(p_{12})(1-p_{32})...(1-p_{\kappa 2})+(1-p_{12})(p_{32})...(1-p_{\kappa 2})+...+\\\nonumber&&(1-p_{12})(1-p_{32})...(p_{\kappa 2})]+\acute{R_1}\cdot(1-p_{12})(1-p_{32})...(1-p_{\kappa  2})\\\nonumber &=&...=\acute{R_2}\cdot[(p_{1\acute{\kappa}})(1-p_{3\acute{\kappa}})...(1-p_{\kappa \acute{\kappa}})+(1-p_{1\acute{\kappa}})(p_{3\acute{\kappa}})...(1-p_{\kappa \acute{\kappa}})\acute{R_2}\\\nonumber&&+...+(1-p_{1\acute{\kappa}})(1-p_{3\acute{\kappa}})...(p_{\kappa \acute{\kappa}})]+\\&&\acute{R_1}\cdot(1-p_{1\acute{\kappa}})(1-p_{3\acute{\kappa}})...(1-p_{\kappa\acute{\kappa}})
  \end{eqnarray}
  
  and for player $A_{\kappa}$
   \begin{eqnarray}
   \label{33}
  \nonumber &&\acute{R_{\kappa}}\cdot(p_{11})(p_{21})...(p_{(\kappa-1) 1})+\acute{R_{\kappa-1}}\cdot[(1-p_{11})(p_{21})...(p_{(\kappa-1) 1})+\\\nonumber&&(p_{11})(1-p_{21})...(p_{(\kappa-1) 1})+(p_{11})(p_{21})...(1-p_{(\kappa-1) 1})]+...+\\\nonumber && \acute{R_2}\cdot [(p_{11})(1-p_{21})...(1-p_{(\kappa-1) 1})+(1-p_{11})(p_{21})...\\\nonumber &&(1-p_{(\kappa-1) 1})+...+(1-p_{11})(1-p_{21})...(p_{(\kappa-1) 1})]+\\\nonumber &&\acute{R_1}\cdot(1-p_{11})(1-p_{21})...(1-p_{(\kappa-1) 1})\\\nonumber &=&\acute{R_{\kappa}}\cdot(p_{12})(p_{22})...(p_{(\kappa-1) 2})+\acute{R_{\kappa-1}}\cdot[(1-p_{12})(p_{22})...(p_{(\kappa-1) 2})\\\nonumber &&+(p_{12})(1-p_{22})...(p_{(\kappa-1) 2})+(p_{12})(p_{22})...(1-p_{(\kappa-1) 2})]\\\nonumber &&+\acute{R_2}\cdot[(p_{12})(1-p_{22})...(1-p_{(\kappa-1) 2})+(1-p_{12})(p_{22})...(1-p_{(\kappa-1) 2})\\\nonumber&&+...+(1-p_{12})(1-p_{22})...(p_{(\kappa-1) 2})]\\\nonumber&&+\acute{R_1}\cdot(1-p_{12})(1-p_{22})...(1-p_{(\kappa-1)  2})\\\nonumber&=&...=\acute{R_{\kappa}}\cdot(p_{1\acute{\kappa}})(p_{2\acute{\kappa}})...(p_{(\kappa-1) \acute{\kappa}})+\acute{R_{\kappa-1}}\cdot[(1-p_{1\acute{\kappa}})(p_{2\acute{\kappa}})...(p_{(\kappa-1) \acute{\kappa}})\\\nonumber&&+(p_{1\acute{\kappa}})(1-p_{2\acute{\kappa}})...(p_{(\kappa-1) \acute{\kappa}})+(p_{11})(p_{21})...(1-p_{(\kappa-1) 1})]\\\nonumber &&+...+\acute{R_2}\cdot[(p_{1\acute{\kappa}})(1-p_{2\acute{\kappa}})...(1-p_{(\kappa-1) \acute{\kappa}})\\\nonumber&&+(1-p_{1\acute{\kappa}})(p_{2\acute{\kappa}})...(1-p_{(\kappa-1) \acute{\kappa}})+...+\\\nonumber &&(1-p_{1\acute{\kappa}})(1-p_{2\acute{\kappa}})...(p_{(\kappa-1) \acute{\kappa}})]+\\&&\acute{R_1}\cdot(1-p_{1\acute{\kappa}})(1-p_{2\acute{\kappa}})...(1-p_{(\kappa-1)\acute{\kappa}})
    \end{eqnarray}
    We will also have,
\begin{equation}
\label{34}
\sum\limits_{i=1}^{\acute{\kappa}}p_{ji}=1;\;\;\;  \forall j.
\end{equation}
Solving above set of equations i.e. (\ref{31},\ref{32},\ref{33},\ref{34}), it turns out that $\forall j,i ; \;\;   p_{ji}=\frac{1}{\acute{\kappa}}$. Applying similar logic as with three player case we can see that it is optimal for every player to randomize on $\kappa$ rounds.

This kind of randomization is only possible if only possible when whitewashing user whitewash in a coordinated. Coordination is required because otherwise the value of $\kappa$ will remain unknown to whitewashing users. If they whitewash in such a way, the whitewashing level will be reduced $\kappa$ times.

It is interesting to note that when all users randomize over $\acute{\kappa}$ rounds, a node can unilaterally increase its pay-off by decreasing the number of rounds, it is going to randomize. It will get maximum pay-off when every other node will randomize over $\acute{\kappa}$ rounds and it will whitewash in every period. As every node is interested in its own pay-off, every node will adapt the strategy to whitewash in every round and hence all will get zero pay-off as shown earlier and consequently they will be discouraged for whitewashing.

\section{Numerical Results}
\begin{figure}[!t]
\begin{center}
\includegraphics[width=80mm, height=85mm, keepaspectratio=false]{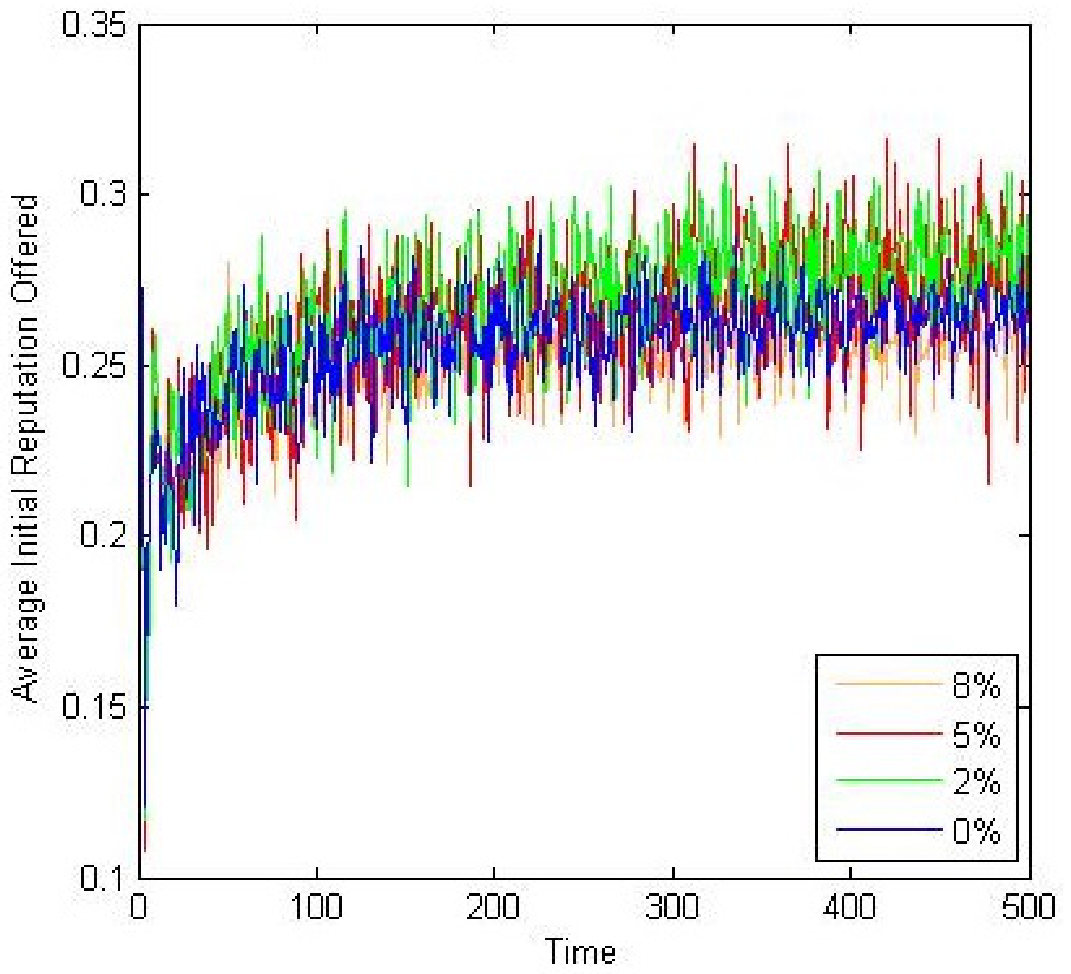}
\caption{Average Initial reputation offered to newcomer nodes}
\label{reppower}
\end{center}
\end{figure}

\begin{figure}[!t]
\begin{center}
\includegraphics[width=80mm, height=85mm, keepaspectratio=false]{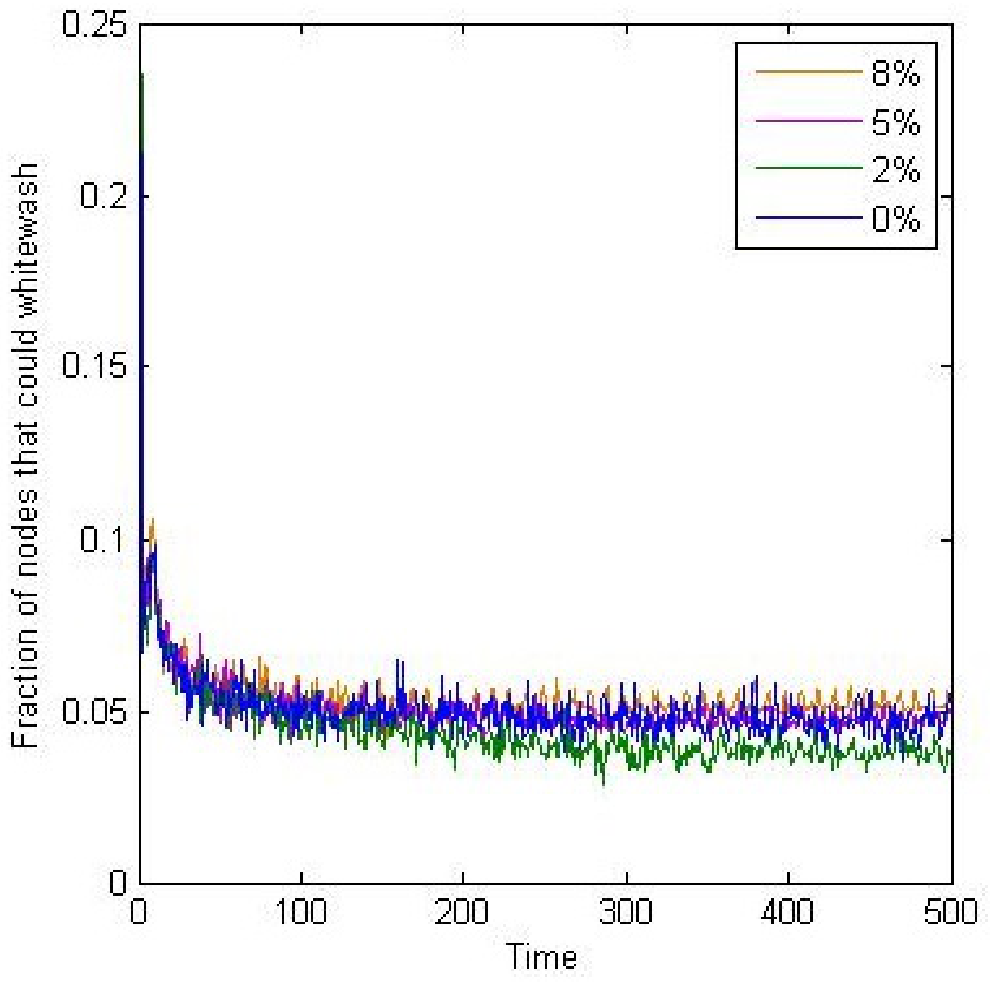}
\caption{Average Initial reputation offered to newcomer nodes}
\label{wwlpower}
\end{center}
\end{figure}

\begin{figure}[!t]
\begin{center}
\includegraphics[width=80mm, height=85mm, keepaspectratio=false]{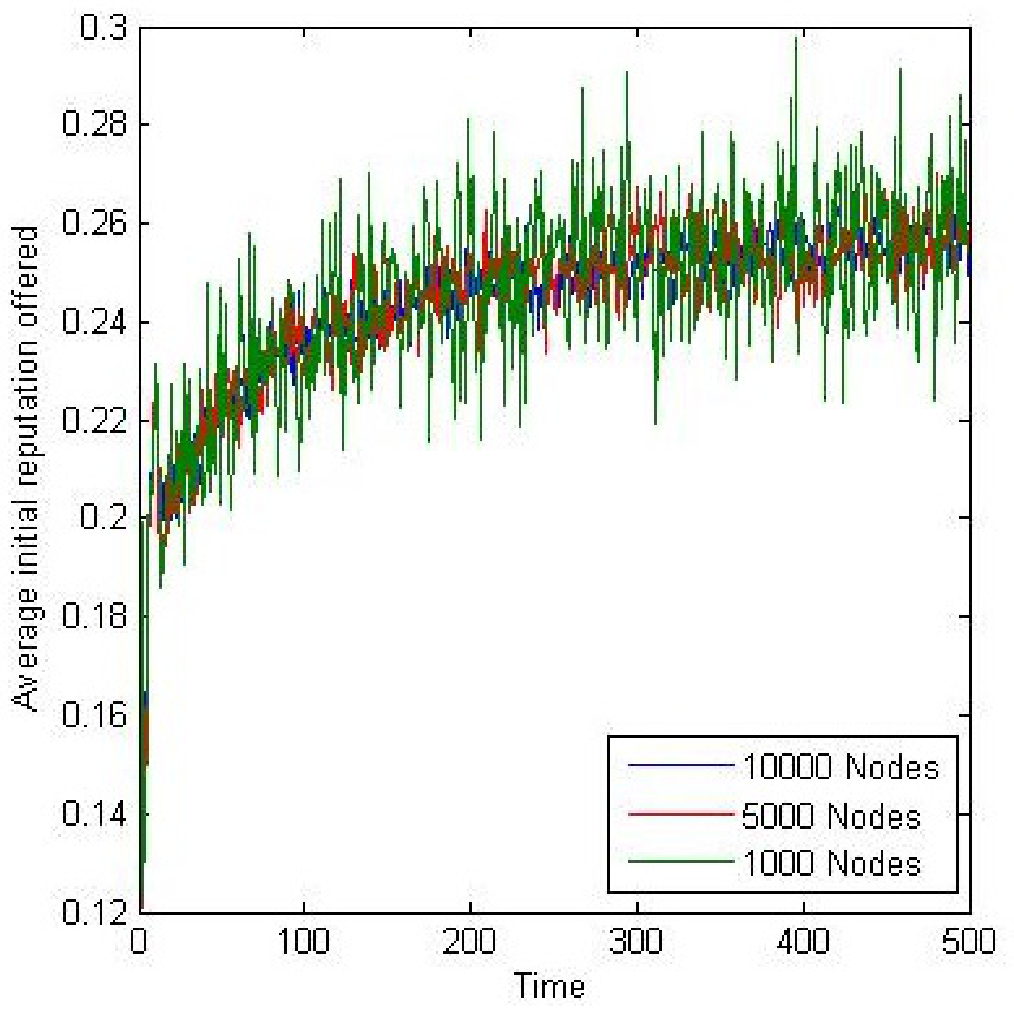}
\caption{Average Initial reputation offered to newcomer nodes}
\label{repreg}
\end{center}
\end{figure}

\begin{figure}[!t]
\begin{center}
\includegraphics[width=80mm, height=85mm, keepaspectratio=false]{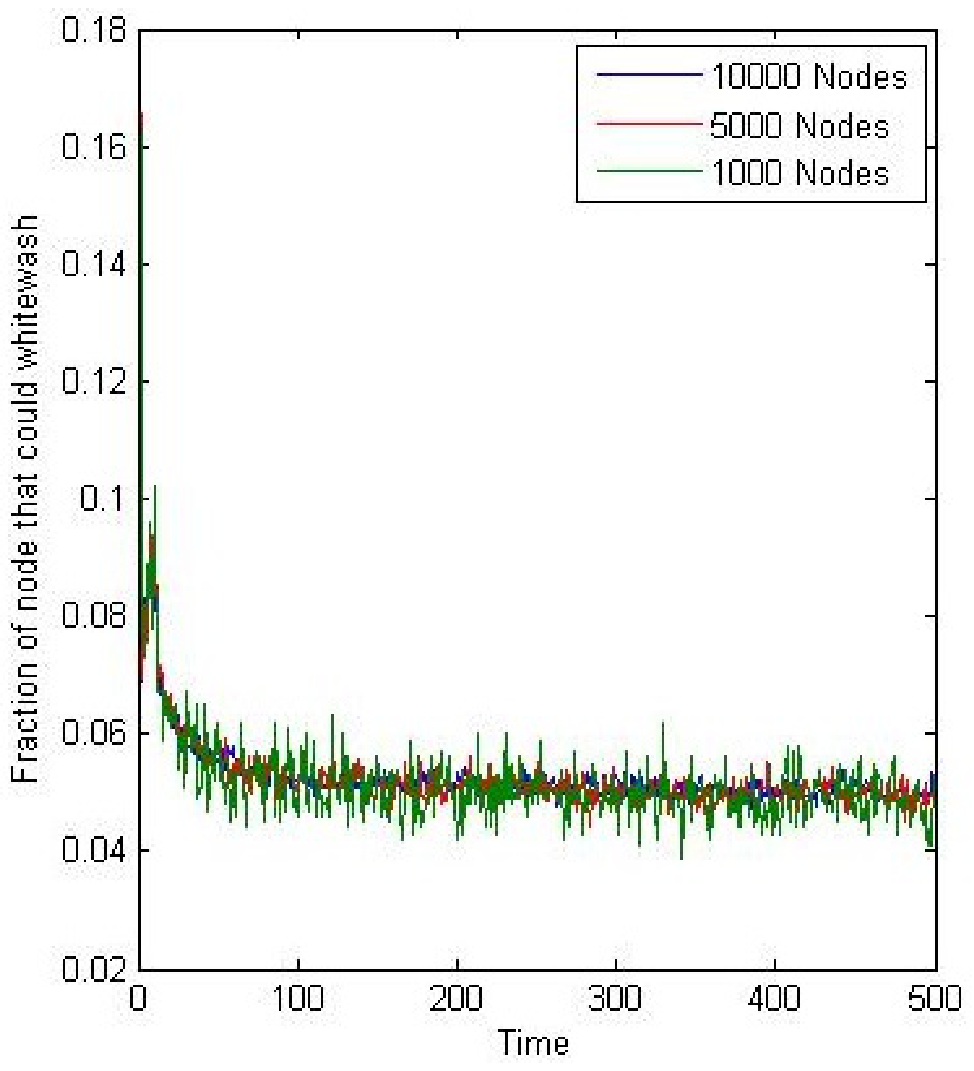}
\caption{Average Initial reputation offered to newcomer nodes}
\label{wwlreg}
\end{center}
\end{figure}

\label{NR4}
Performance evaluation of the proposed method to avoid whitewash has been done for different kinds of networks, viz. growing scale-free networks based on PA model starting with 1000 nodes with growth rate of 0\% per 10 iteration, 2\% per 10 iteration, 5\% per 10 iteration, 8\% per 10 iteration and regular network with $1000$, $5000$ and $10000$ nodes. By growing network, we mean in such a network after every 10 iterations some percentage of new nodes join the network. 

We have considered the discrete time instants for the purpose of measurement and estimation in the simulations. Every slot is termed as an iteration. Value of $R_{ini,i,max}$ and $R_{ini,i,min}$ for all nodes has been taken as $0.5$ and $0.03$ respectively. Thus, 50\% of the nodes are taken to be potential whitewasher initially. 

In the first iteration, all potential whitewasher nodes attempt to whitewash to randomly chosen nodes. Every node calculates the level of whitewash and according to that it offers the initial reputation. Nodes, that attempt for whitewash, go for it as per their honesty level and offered initial reputation. From next round onwards, potential whitewashing nodes attempt for whitewash probabilistically and the probability for attempting whitewash for a node depends on the number of times that node is able to whitewash, i.e.,
\begin{equation}
P_{wa}=\frac{number\; of\; times\; nodes\; was\; able\; to\; whiewash}{total\; number\; of\; times\; node\; attempted\; for\; whitewash}.
\end{equation}
Here $P_{wa}$ is the probability by which a node will attempt whitewash. This process is repeated again and again up to 500 iteration.

Figure \ref{wwlpower} and figure \ref{wwlreg} present the fraction of nodes that could whitewash in every iteration for growing scale-free network and regular networks respectively. 
Figure \ref{reppower} and figure \ref{repreg} present the average initial reputation offered by nodes in every iteration for growing scale-free network and regular networks respectively.  

It is evident in figures that after some iterations, very few nodes could whitewash in spite of a reasonable initial reputation being offered to newcomer nodes. This can also be observed from figures that proposed method works for different kinds of networks.
\begin{figure}[!t]
\begin{center}
\subfigure[]{
\includegraphics[width=65mm, height=55mm, keepaspectratio=false]{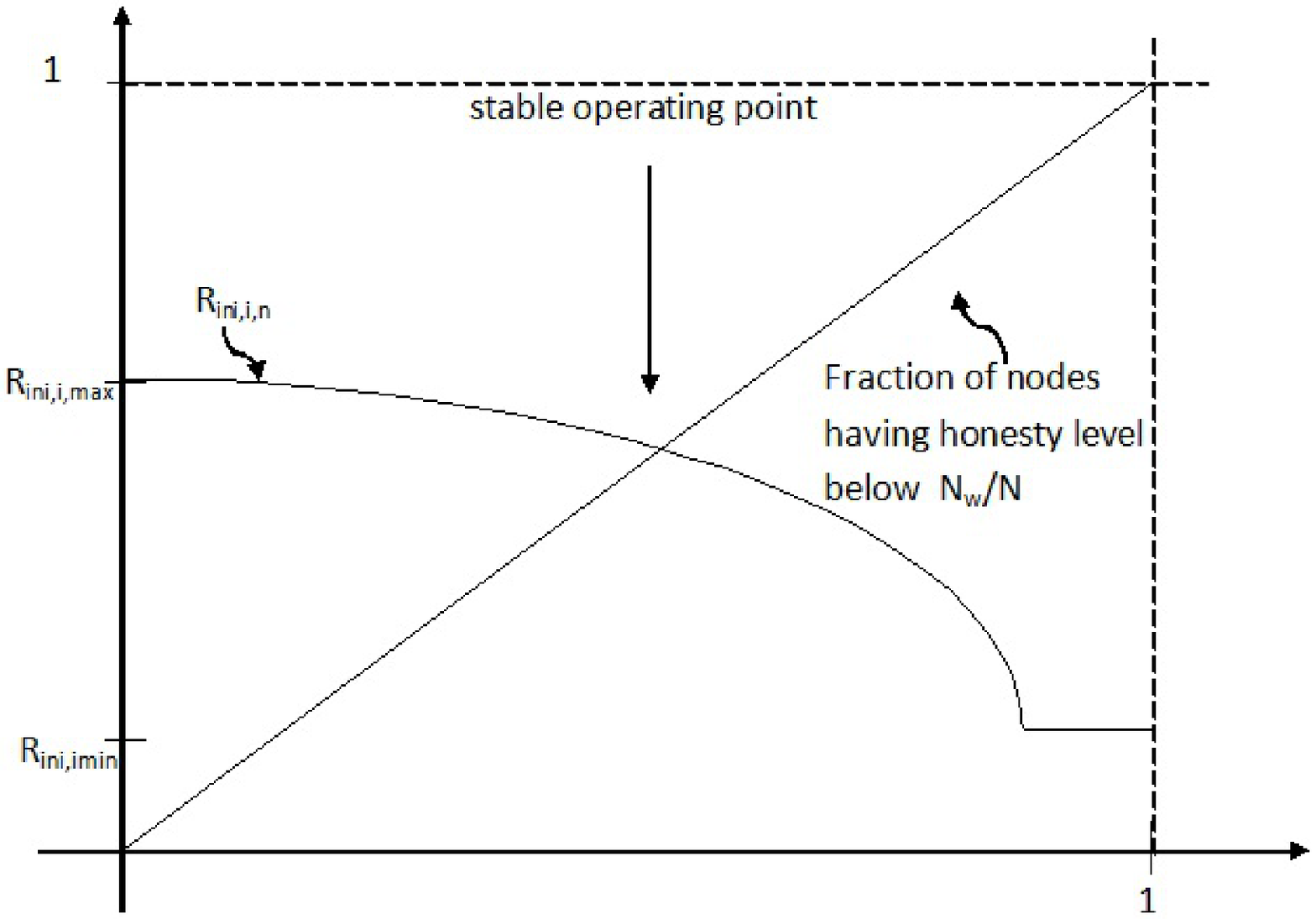}
\label{wwca}
}
\subfigure[]{
\includegraphics[width=65mm, height=55mm, keepaspectratio=false]{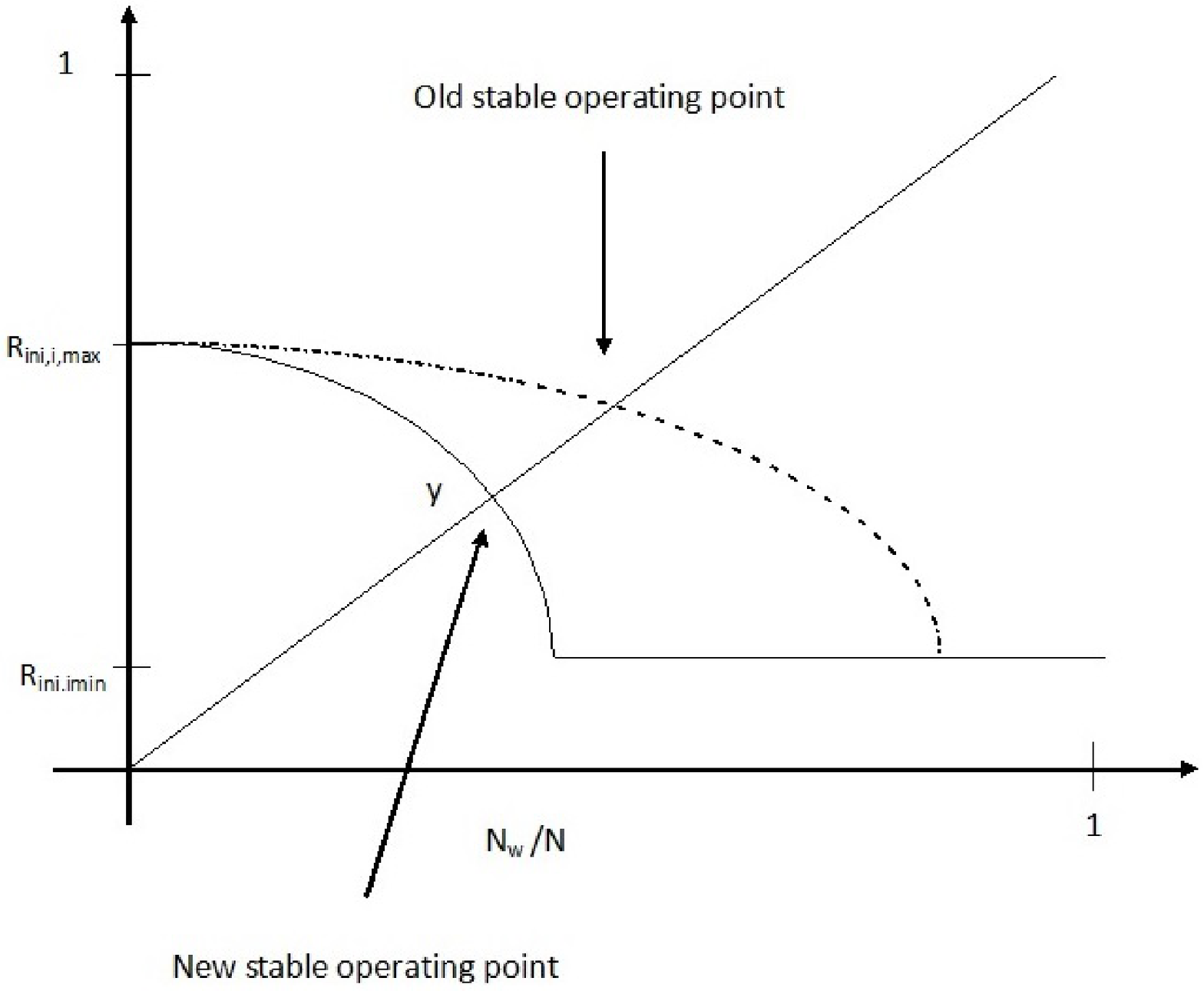}
\label{wwcb}
}
\caption{Change in Maximum whitewashing level}
\label{wwc}
\end{center}
\end{figure}
In figure \ref{wwc}, we have plotted simultaneously, the fraction of users having honesty level below $\frac{N_w}{N}$ and initial reputation as function of whitewashing level $\frac{N_w}{N}$. The intersection of these two graphs gives the stable operating point.

Decreasing level of whitewash can be explained by the fact that, as our algorithm periodically estimates maximum whitewashing level, once system attains stability, the minimum reputation whitewashing level will be reduced. This gives new stable operating point Y (as shown in figure \ref{wwcb}) point. This process will be repeated again and again and over the time, the operating point adjusted to keep the whitewashing level negligible value.

\section{Conclusion}
\label{CON4}
Whitewashing tendency of selfish users is a big problem in implementation of reputation management system in peer-to-peer networks. In this paper, we have proposed a method to avoid the problem of whitewashing. This method adjusts initial reputation value as per the level of whitewash. It is proved theoretically and by simulation that the proposed method is able to avoid whitewash \nocite{ruchirestm}. 

\bibliography{ref}{}
\bibliographystyle{ieeetr}
\end{document}